# Groverian Entanglement Measure and Evolution of Entanglement in Search Algorithm for n(=3, 5)-Qubit Systems with Real Coefficients.

**Arti Chamoli and C. M. Bhandari**
**Indian Institute of Information Technology Allahabad, Deoghat, Jhalwa, Allahabad-211011, India**
Email: achamoli@iiita.ac.in, cmbhandari@yahoo.com

Evolution of entanglement with the processing of quantum algorithms affects the outcome of the algorithm. Particularly, the performance of Grover's search algorithm gets worsened if the initial state of the algorithm is an entangled one. The success probability of search can be seen as an operational measure of entanglement. This paper demonstrates an entanglement measure based on the performance of Grover's search algorithm for three and five qubit systems. We also show that although the overall pattern shows growth of entanglement, its rise to a maximum and then consequent decay, the presence of local fluctuation within each iterative step is likely.

**PACS**
03.67.-a, 03.65.Ud.

## INTRODUCTION

Quantum entanglement [1] and superposition [2] are the pillars of quantum computation and quantum information theory [3,4]. Quantum information theory has reached the new arenas exploiting these two. Quantum entanglement, inherently a non-classical phenomenon, signifies correlations between quantum systems even if they are space-like separated. In recent times it has been reckoned as a physical resource and hence utilized for various computational tasks including quantum information processing [5] and cryptography [6]. Hence, quantification of entanglement of quantum states attains utmost importance. Various entanglement evaluating measures have been figured by current researchers. Methodology based on operational considerations have been successfully employed to formulate entanglement measures for bipartite systems [7,8]. Based on correspondence between thermodynamics and entanglement, entropy of entanglement has been considered as the unique measure of entanglement of pure states [9]. Applications like quantum teleportation can only be materialized if certain amount of entanglement exists between the communicators initially. A successful teleportation requires the two parties to share particles in maximally entangled states. In the absence of maximally entangled pairs one of the parties can perform local operations to concentrate the entanglement leading to creation of *n* maximally entangled pairs out of *k* non-maximally entangled pairs, where n < k. For large k, the ratio $n/k$ asymptotically leads to entropy of entanglement. In addition to this, certain attributes have been framed to measure entanglement [10-13]. These attributes are based on axiomatic considerations. According to these, any entanglement should not prevail in product states, it should not vary with local unitary operations and should not increase consequential to any sequence of local operations complemented by only classical communication between parties. Measures satisfying the above properties are called entanglement monotones [11].

In quantum search algorithm entanglement is not looked upon as a physical resource. On the contrary the starting state best suited is a product state with uniform superposition. However, it is interesting and illustrative to see that even though it is not explicitly sought for, entanglement invariably creeps in as the search progresses. It is thus possible to link the success rate of search algorithm with an operational entanglement measure. Based on these considerations an entanglement measure has been developed by Biham et al [14]. The measure is based on the linkage of the success of Grover's search algorithm [15,16] to the amount of entanglement present in the initial state. Performance of Grover's algorithm deteriorates with increasing entanglement in the initial state. Considering the modified quantum search as given in [14] in which a product of arbitrary local operations is applied to intial input register, the formulation of maximal probability of success, $P_{\max}(\Psi)$, as an entanglement monotone can be precisely made. For a search space containing $N=2^n$ elements, where n is an integer, the elements can be represented by an n-qubit register and the intial register as $|\Phi\rangle$. For a single marked solution, $s$, to the search problem, $P_{\max}$ in terms of the operator $U_G^m$, representing $m$ Grover iterations may be written as

$$P_{\max} = \max_{U_1,\ldots U_n} \frac{1}{N} \sum_{s=0}^{N-1} \left| \langle s | U_G^m (U_1 \otimes U_2 \otimes \ldots U_n) | \Phi \rangle \right|^2$$

(1)

by averaging uniformly over all $N$ possible values for $s$. The maximization is over all local unitary operations $U_1,\ldots U_n$ on the respective qubits of input register state $|\Phi\rangle$. This can be generalized by considering the action of the Grover iterations on the equal superposition state $|\eta\rangle = \frac{1}{\sqrt{N}} \sum_x |x\rangle$. Applying in Grover iterations yields

$$U_G^m |\eta\rangle = |s\rangle + O\left(\frac{1}{\sqrt{N}}\right)$$

(2)

where the second term is a small correction because Grover's algorithm yields a solution with probability $1 - O\left(\frac{1}{\sqrt{N}}\right)$. Multiplying eq. (2) by $\left(U_G^m\right)^\dagger$ and then taking the Hermitian conjugate gives

$$\langle s | U_G^m = \langle \eta | + O\left(\frac{1}{\sqrt{N}}\right)$$

(3)

Substituting in eq. (1) gives, for a general state $|\Phi\rangle$,

$$P_{\max} = \max_{U_1,\ldots U_n} \frac{1}{N} \sum_{s=0}^{N-1} \left| \langle \eta | U_G^m (U_1 \otimes U_2 \otimes \ldots U_n) | \Phi \rangle \right|^2 + O\left(\frac{1}{\sqrt{N}}\right)$$

(4)

Since $|\eta\rangle$ is a product state, eq.(4) may equivalently may be expressed as

$$P_{max} = \max_{|e_1,....e_n\rangle} |\langle e_1,....e_n|\Phi\rangle|^2 + O\left(\frac{1}{\sqrt{N}}\right)$$

(5)

where the maximization now runs over all product states, $|e_1,....e_n\rangle = |e_1\rangle \otimes ...... \otimes |e_n\rangle$, of the $n$ qubits. This suggests that $P_{max}$ depends on the maximum of the overlap between all product states and the input state $|\Phi\rangle$. For a product state as input state, $P_{max}$ would be equal to one, whereas with an entangled state as input state, $P_{max}$ would never be one. Success probability of the search algorithm depends on the entanglement of initial register state. Quantifying entanglement following the above approach is related to the performance of the quantum state as an input to the modified search algorithm. The measure thusly referred to as Groverian entanglement can be defined for a state $|\Psi\rangle$ by

$$G(\Psi) = \sqrt{1 - P_{max}}$$

(6)

$P_{max}(\Psi)$ is an entanglement monotone and consequently $G(\Psi)$ too. Following the same line of reasoning authors [17] have examined the success rate of Grover's search algorithm for various four qubit states and Groverian entanglement measure has been worked out for certain kind of input states. In this letter, the authors have evaluated the success rate of Grover's search algorithm for three and five qubit states and Groverian entanglement measure has been formulated for the same.

**THREE-QUBIT STATES**

An arbitrary initial state $|\Psi\rangle$ of three qubits can be written as

$$|\Psi\rangle = \sum_{i=0}^{7} a_i |i\rangle$$

(7)

where $|i\rangle = |i_0 i_1 i_2\rangle$.

It can have upto eight terms. Now a general product state of three qubits can be written as

$$|e\rangle = |e_1\rangle \otimes |e_2\rangle \otimes |e_3\rangle$$

(8)

where a single qubit can be represented as

$$|e_k\rangle = \cos\theta_k |0\rangle_k + e^{i\Phi_k} \sin\theta_k |1\rangle_k$$

(9)

where $k = 1,2,3$. Angle $\theta_k$ is in the range $0 \leq \theta_k \leq \pi/2$, while $\Phi_k$ is in the range $0 \leq \Phi_k \leq 2\pi$.

The Groverian entanglement measure is derived through the maximization of the function,

$$P(\theta_1,\theta_2,\theta_3,\Phi_1,\Phi_2,\Phi_3,\Psi) = |\langle e|\Psi\rangle|^2$$

(10)

with respect to variables $\theta_k, \Phi_k$ (where $k = 1, 2, 3$) and $\Psi$.

The product state $|e\rangle$ has supposedly real amplitudes only with all $\Phi_k = 0$ or $\pi$ if initial state is the one for which all $a_i$'s are real. In order to discuss the success probability of this particular case, the maximization over $\Phi_k$ is reduced to a discrete maximization with $e^{i\Phi_k} = \pm 1$. Doubling the range of $\theta_k$ to $-\pi/2 \leq \theta_k \leq \pi/2$ makes $\sin\theta_k$ to be both positive and negative for same value of $\cos\theta_k$ thereby neutralizing the presence of $i\Phi_k$. Hence, $\Psi$ with real $a_i$'s will have its expression for maximum success probability as

$$P_{max}(\Psi) = \max_{\theta_1,\theta_2,\theta_3} P(\theta_1,\theta_2,\theta_3,\Psi)$$

(11)

The function P is given by

$$P(\theta_1,\theta_2,\theta_3,\Psi) = [a_{000}\cos\theta_1\cos\theta_2\cos\theta_3 + a_{001}\cos\theta_1\cos\theta_2\sin\theta_3 + a_{010}\cos\theta_1\sin\theta_2\cos\theta_3 + a_{011}\cos\theta_1\sin\theta_2\sin\theta_3 + a_{100}\sin\theta_1\cos\theta_2\cos\theta_3 + a_{101}\sin\theta_1\cos\theta_2\sin\theta_3 + a_{110}\sin\theta_1\sin\theta_2\cos\theta_3 + a_{111}\sin\theta_1\sin\theta_2\sin\theta_3]^2$$

(12)

With the help of trigonometric identities, expression for P can written as,

$$P(\theta_w, \theta_x, \theta_y, \theta_z, \Psi) = [\frac{a_{000}-a_{110}-a_{101}-a_{011}}{4}\cos\theta_w + \frac{a_{100}+a_{010}+a_{001}-a_{111}}{4}\sin\theta_w + \frac{a_{000}-a_{110}+a_{101}+a_{011}}{4}\cos\theta_x + \frac{a_{100}+a_{010}-a_{001}+a_{111}}{4}\sin\theta_x + \frac{a_{000}+a_{110}-a_{101}+a_{011}}{4}\cos\theta_y + \frac{a_{100}-a_{010}+a_{001}+a_{111}}{4}\sin\theta_y + \frac{a_{000}+a_{110}+a_{101}-a_{011}}{4}\cos\theta_z + \frac{a_{100}-a_{010}-a_{001}-a_{111}}{4}\sin\theta_z]^2$$

(13)

where $\theta_w = \theta_1 + \theta_2 + \theta_3$, $\theta_x = \theta_1 + \theta_2 - \theta_3$, $\theta_y = \theta_1 - \theta_2 + \theta_3$, $\theta_z = \theta_1 - \theta_2 - \theta_3$.

Maximization of P is obtained by maximizing $P(\theta_w, \theta_x, \theta_y, \theta_z, \Psi)$ with respect to $\theta_w$, $\theta_x$, $\theta_y$, and $\theta_z$. $P_{max}(\Psi)$ is thus obtained by satisfying the condition of maxima for $P(\theta_w, \theta_x, \theta_y, \theta_z, \Psi)$,

$$\text{i.e.,} \frac{\partial P}{\partial \theta_w} = \frac{\partial P}{\partial \theta_x} = \frac{\partial P}{\partial \theta_y} = \frac{\partial P}{\partial \theta_z} = 0$$

(14)

This leads to the following expression for maximum success probability for three qubits case.

$$P_{max} = \frac{1}{16}\left[\sqrt{\{(a_{000} - a_{110} - a_{101} - a_{011})^2 + (a_{100} + a_{010} + a_{001} - a_{111})^2\}} + \right.$$
$$\sqrt{\{(a_{000} - a_{110} + a_{101} + a_{011})^2 + (a_{100} + a_{010} - a_{001} + a_{111})^2\}} +$$
$$\sqrt{\{(a_{000} + a_{110} - a_{101} + a_{011})^2 + (a_{100} - a_{010} + a_{001} + a_{111})^2\}} +$$
$$\left.\sqrt{\{(a_{000} + a_{110} + a_{101} - a_{011})^2 + (a_{100} - a_{010} - a_{001} - a_{111})^2\}}\right]^2$$

(15)

With the analytical expression for three qubits, $P_{max}(\Psi)$ can be calculated for various choices of $|\Psi\rangle$. For a product state, the entanglement is zero, this can be easily verified. For example uniform product state can be obtained with three qubits in either of these states:

$$\frac{1}{\sqrt{2}}(|0\rangle + |1\rangle), \frac{1}{\sqrt{2}}(|0\rangle - |1\rangle)$$

The substitution of $a_i$'s in the analytical expression gives $P_{max}(\Psi) = 1$, and thereby giving $G(\Psi) = 0$.

Eq. (15) makes it possible to evaluate the value of Groverian entanglement measure for a general three qubit state with real coefficients. Some such cases will be discussed.

As discussed earlier (Biham et al, [14]), the search algorithm starts with a product state with zero entanglement, it evolves as the iterative operation proceeds, reaching a maximum, and then decays. It would be of interest to describe the evolution of entanglement for three qubit system. If we consider the state of uniform superpositon which is a linear combination of eight terms

$$|\Psi\rangle = \frac{1}{2\sqrt{2}}(|000\rangle + |001\rangle + |010\rangle + |011\rangle + |100\rangle + |101\rangle + |110\rangle + |111\rangle)$$

and find out the amount of entanglement present in each intermediate state then it is observed that entanglement grows, reaches its maximum and eventually fades away as we reach the desired state. The intermediate states can be obtained by applying the operators $P_w$ and $P_s$ successively to the initial state. $P_w$ is an operator of the form $1 - 2|w\rangle\langle w|$, where $|w\rangle$ is the desired state. $P_s$ is the operator of the form $1 - 2|\Psi\rangle\langle\Psi|$.

The Groverian entanglement measure $G(\Psi)$ as a function of the no. of iterations has been plotted in Fig. 1. It is of interest to note the change of $G(\Psi)$ within each iterative step: the first step (operator $P_w$) raises $G(\Psi)$ to around 0.37 and second step (operator $P_s$) reduces it to 0.26. A similar variation is seen in the second iterative step where $G(\Psi)$ rises to 0.28 and then finally drops to 0.14. Expectedly the entanglement should approach zero with $P_{max}$ approaching 1. However, it is not usually so, as each iterative operation rotates the state vector in the direction of the desired state with a certain angle *θ*, depending upon number of qubits, *n*, which at times rotates the initial state exactly to the desired state [18].

**FIVE-QUBIT STATES**
Following the same procedure as earlier, an analytical expression for maximum success probability of Grover's search algorithm for five qubit states has been obtained. The expression is given in the Appendix. A general five qubits state is a linear combination of thirty two terms. Groverian entanglement measure for the same can be calculated. For example, the general product state $|\Psi\rangle$ with uniform superposition is of the form

$$|\Psi\rangle = \sum_{i=0}^{31} a_i |i\rangle, \text{ where } |i\rangle = |i_0 \ldots i_4\rangle \text{ and } a_i = \frac{1}{4\sqrt{2}} \text{ for all i.}$$

On substituting the value of coefficients of each state it can be seen that $P_{max}(\Psi) = 1$, and thus $G(\Psi) = 0$, verifying that a product state has zero entanglement.

Thus maximum success probability and hence Groverian entanglement can be obtained for certain five qubit states from the analytical expression given in the appendix.

Once again it can be shown for a five qubit state that success probability of Grover's search algorithm is affected by the entanglement present in the initial state. Entanglement limits the success probability. Also it can be verified that entanglement grows and then dies out as the iterative operation of the search algorithm goes forward. Fig. 2 displays Groverian entanglement as a function of the no. of iterations of search algorithm. The pattern of variation within each iterative step is non-monotonic and similar to one displayed in fig.1.

**OTHER ENTANGLEMENT MEASURES**

**Von Neumann entropy as an entanglement measure**
We start with two observers, Alice and Bob, who share *k* pairs in non-maximally entangled state. By local filtering operations it is possible to produce *n* maximally entangled pairs. This transformation is reversible for large value of *n* and *k* and can be done by either of the observers by performing local operations. The ratio $n/k$ tends to some constant value as *k* tends to infinity. Thus we can have a common entanglement measure for both k and n pairs. The measure of entanglement for *k* systems approaches the entanglement measure for *n* singlets,

$$kE = n$$

where $E$ is the entanglement measure. For singlets entanglement is 1. Hence,

$$E = \lim_{n,k \to \infty} \frac{n}{k}$$

(16)

Bennett et al. [7] have shown that this limit is equal to entropy of entanglement (von Neumann entropy) of the $k$ type non-maximally entangled pairs.

For a system composed of subsystems A and B, the von Neumann entropy of the reduced density matrix of sub system A is given by

$$S(\rho_A) = -\text{Tr}[\rho_A \log \rho_A]$$

(17)

The entanglement of a quantum system can be quantified as bipartite entanglement by calculating its entropy of entanglement, which is expressed as the von Neumann entropy of the reduced state of one of its subsystem. However, for a quantum system with n > 2 subsystems, this quantification cannot be done precisely. In [19], a quantum system with any arbitrary number of subsystems has been considered as bipartite, with one subsystem consisting of a single qubit and the second subsystem all the rest. The reduced density matrix can be calculated for any single qubit because von Neumann entropy is independent of the choice of remaining qubits. The reduced density matrix for the $l$th qubit can be written in its standard form as,

$$\rho_l(k) = \frac{1}{2}[I + \vec{s}(k) \cdot \vec{\sigma}]$$

(18)

where components of the Bloch vector $\vec{s}(k)$, for the intermediate states of Grover's search algorithm can be calculated. The state after $k$ Grover iterations becomes

$$|\Psi_k\rangle = \frac{\cos \theta_k}{\sqrt{N-1}} \sum_{x \neq y} |x\rangle + \sin \theta_k |y\rangle$$

(19)

where $\theta_k = (2k+1)\theta_0$ and $\theta_0$ satisfies $\sin \theta_0 = 1/\sqrt{N}$, and $N = 2^n$ for an $n$ qubit system. Hence after $k$ iterations, the components of Bloch vector $\vec{s}(k)$ are

$$s_x(k) = \frac{N-2}{N-1} \cos^2 \theta_k + \frac{1}{\sqrt{N-1}} \sin 2\theta_k$$

$$s_y(k) = 0$$

$$s_z(k) = \frac{1}{N-1} \cos^2 \theta_k - \sin^2 \theta_k$$

(20)

the von Neumann entropy can thus be calculated as

$$S[\rho_l(k)] = -Tr[\rho_l(k) \log \rho_l(k)]$$

(21)
Following the same approach as in [19], entropy of entanglement has been calculated for intermediate states of Grover's search algorithm for three and five qubit systems. The results have been shown in table 1. It is quite conspicuous from figures 1 and 2 that degree of entanglement as calculated from Groverian entanglement measure and entropy of entanglement as well, for intermediate states of search algorithm follow the same pattern. The presence of local maxima after every $P_w$ rotation can be observed by both the measures. Thus both these measures support the fact that intermediate states of search algorithm through which the system evolves are entangled. Despite of initial and target states being the product states. Due to lack of any precise measure of entanglement for quantum systems with n > 2 subsystems, it is difficult at the moment to say which of the two measures is more appropriate for measuring actual amount of entanglement present.

**Three way tangle**
Entanglement between a pure state of three qubits can be evaluated as residual tangle [20]. For a pure state of three qubits X, Y, and Z, the tangle between any pair of qubits is same. Considering a pair of qubits as a single entity, tangle between any such pair and the remaining qubit of the tripartite system can also be evaluated likewise. Thus the entanglement in a tripartite system can be formulated in terms of pair wise entanglements, as residual tangle or three – way – tangle, $T\{XYZ\}$,

$$T\{XYZ\} = T\{X(YZ)\} - T\{XY\} - T\{XZ\}$$

(22)

where $T\{XY\}$ is the entanglement between X and Y-qubits, calculated from density matrix, $\rho\{XY\}$, that can be calculated by tracing out qubit Z. Similarly, $T\{XZ\}$ is the entanglement between X and Z-qubits and $T\{X(YZ)\}$ is the entanglement between X and YZ (pair).

Eq. (22) can be expressed in terms of the coefficients of eight basis vectors of a three qubit system [20].

$$T\{XYZ\} = 4|d_1 - 2d_2 + 4d_3|$$

(23)

where
$$d_1 = a_{000}^2 a_{111}^2 + a_{001}^2 a_{110}^2 + a_{010}^2 a_{101}^2 + a_{100}^2 a_{011}^2$$
$$d_2 = a_{000}a_{111}a_{011}a_{100} + a_{000}a_{111}a_{101}a_{010} + a_{000}a_{111}a_{110}a_{001}$$
$$\quad + a_{011}a_{100}a_{101}a_{010} + a_{011}a_{100}a_{110}a_{001} + a_{101}a_{010}a_{110}a_{001}$$
$$d_3 = a_{000}a_{110}a_{101}a_{011} + a_{111}a_{001}a_{010}a_{100}$$

Using eq. (23) entanglement within the intermediate states of Grover's search algorithm for three qubit system has been calculated and plotted, as shown in fig.1.

**IN CONCLUSION**

Analytical expressions for Groverian entanglement measure have been obtained for three and five -qubit states. Four qubit case has been described earlier[17]. The measure is a direct consequence of maximum success probability of Grover's search algorithm. Entanglement can be calculated for any state by considering it as the initial state of search algorithm. The fact that Grover's search algorithm performs inaccurately for entangled states has been exploited here. An analytical expression has its own benefits as the amount of entanglement can be figured out for varied choices of initial states ranging from a linear combination of two basis vectors to maximum no. of basis vectors pertaining to a given state.

Evolution of entanglement during the iterative procedure follows a certain pattern. It rises, reaches a maximum and then decays to zero as the desired state is reached. However, the changes are not monotonic and one clearly notices the pattern of variation within each iterative step. Application of operation $P_w$ on $|\Psi\rangle$ tends to increment the value of $G(|\Psi\rangle)$ and application of $P_s$ on the corresponding state lowers it.

A comparison of Groverian measure with entropy of entanglement [7] and three-way-tangle [20] has been shown in table 1 and table 2. Although the calculated value of entanglement for each state within the search process comes out to be different by the three entanglement measures but the entanglement evolves in the similar fashion within the iterative steps as shown in fig.1 and fig.2. The non-monotonic variation is due to division of each iterative step into two parts. However if $G(\Psi)$ is plotted against integral number of steps defined by $(P_S P_W)$ then the variation will be monotonic.

Table 1.
Evolution entanglement within three qubits search algorthim as calculated by Groverian Entanglement Measure, Entropy of Entanglement, Three-way Tangle.

| Starting State | Groverian Entanglement Measure | Entropy of Entanglement | Three-way Tangle |
|---|---|---|---|
| $|\Psi\rangle$ (uniform superposition) | 0 | 0.08 | 0 |
| $P_W|\Psi\rangle$ | 0.38 | 0.84 | 0.25 |
| $P_S P_W|\Psi\rangle$ (state after 1st iteration) | 0.27 | 0.31 | 0.0625 |
| $P_W P_S P_W|\Psi\rangle$ | 0.29 | 0.54 | 0.1406 |
| $P_S P_W P_S P_W|\Psi\rangle$ (state after 2nd iteration) | 0.15 | 0.19 | 0.0224 |

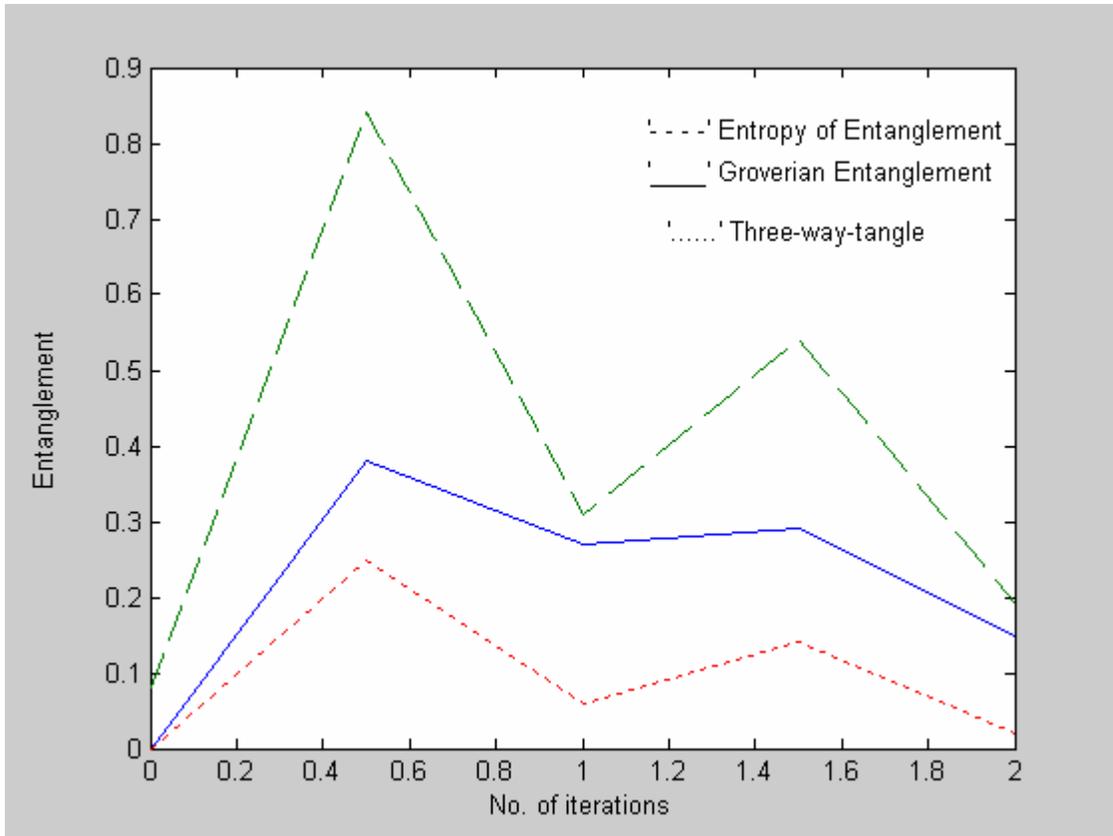

Figure1. Various entanglement measures as a function of the no. of iterations for three qubit quantum search algorithm. The dashed, solid and dotted lines represent evolution of entanglement as given by entropy of entanglement, Groverian measure and three-way-tangle respectively.

Table2.
Evolution entanglement within five qubits search algorithm as calculated by Groverian Entanglement Measure, Entropy of Entanglement, Three-way Tangle.

| Starting State | Groverian Entanglement Measure | Entropy of Entanglement |
|---|---|---|
| $\lvert\Psi\rangle$ (uniform superposition) | 0 | 0.14 |
| $P_W\lvert\Psi\rangle$ | 0.24 | 0.39 |
| $P_S P_W\lvert\Psi\rangle$ (state after 1$^{st}$ iteration) | 0.23 | 0.31 |
| $P_W P_S P_W\lvert\Psi\rangle$ | 0.57 | 0.49 |
| $P_S P_W P_S P_W\lvert\Psi\rangle$ (state after 2$^{nd}$ iteration) | 0.34 | 0.47 |
| $P_W P_S P_W P_S P_W\lvert\Psi\rangle$ | 0.38 | 0.49 |
| $P_S P_W P_S P_W P_S P_W\lvert\Psi\rangle$ (state after 3$^{rd}$ iteration) | 0.22 | 0.25 |
| $P_W P_S P_W P_S P_W P_S P_W\lvert\Psi\rangle$ | 0.21 | 0.31 |
| $P_S P_W P_S P_W P_S P_W P_S P_W\lvert\Psi\rangle$ (state after 2nd iteration) | 0.03 | 0 |

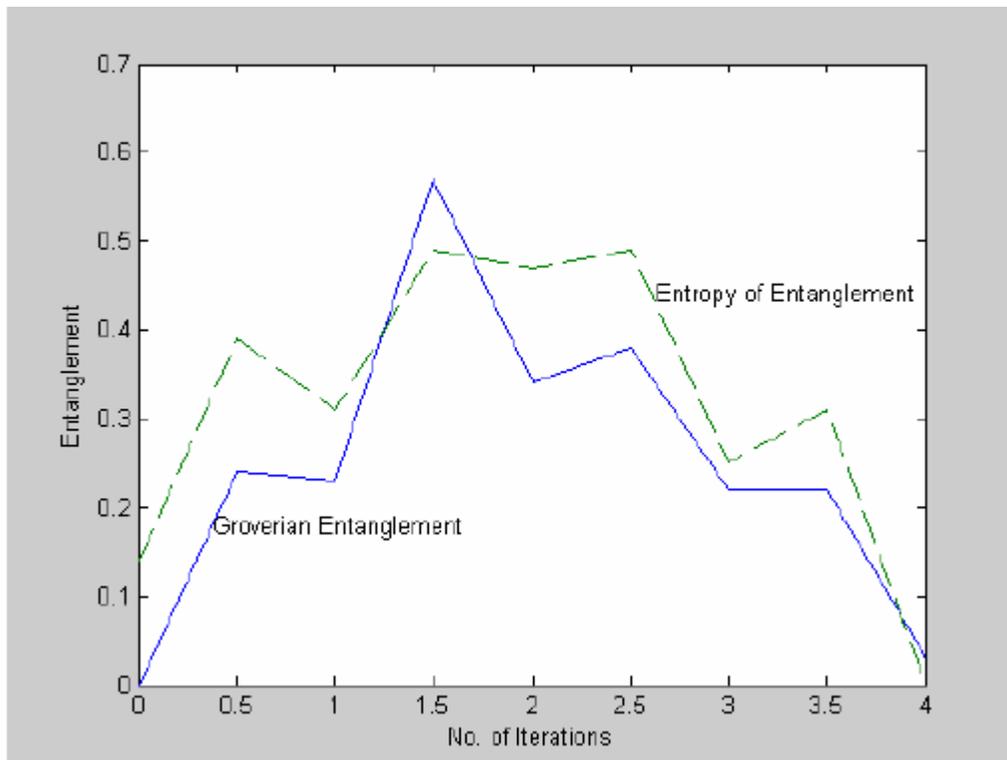

Figure2. Entanglement as a function of the no. of iterations of search algorithm for five qubit system as given by entropy of entanglement and Groverian measure.

**ACKNOWLEDGEMENT**
Authors are thankful to Dr. M. D. Tiwari for his keen interest and support. Arti Chamoli is thankful to IIIT, Allahabad for financial support.

## Appendix

Expression for success probability of Five-Qubit States

$$P_{max} = \frac{1}{256}[1\sqrt{(a_{00000} - a_{11000} - a_{10100} - a_{01100} - a_{10010} - a_{01010} - a_{00110} + a_{11110} - a_{10001} - a_{01001} - a_{00101} + a_{11101} - a_{00011} + a_{11011} + a_{10111} + a_{01111})^2 + (a_{10000} + a_{01000} + a_{00100} - a_{11100} + a_{00010} - a_{11010} - a_{10110} - a_{01110} + a_{00001} - a_{11001} - a_{10101} - a_{01101} - a_{10011} - a_{01011} - a_{00111} + a_{11111})^2} +$$

$$\sqrt{(a_{00000} - a_{11000} - a_{10100} - a_{01100} - a_{10010} - a_{01010} - a_{00110} + a_{11110} + a_{10001} + a_{01001} + a_{00101} - a_{11101} + a_{00011} - a_{11011} - a_{10111} - a_{01111})^2 + (a_{10000} + a_{01000} + a_{00100} - a_{11100} + a_{00010} - a_{11010} - a_{10110} - a_{01110} - a_{00001} + a_{11001} + a_{10101} + a_{01101} + a_{10011} + a_{01011} + a_{00111} - a_{11111})^2} +$$

$$\sqrt{(a_{00000} - a_{11000} - a_{10100} - a_{01100} + a_{10010} + a_{01010} + a_{00110} - a_{11110} - a_{10001} - a_{01001} - a_{00101} + a_{11101} + a_{00011} - a_{11011} - a_{10111} - a_{01111})^2 + (a_{10000} + a_{01000} + a_{00100} - a_{11100} - a_{00010} + a_{11010} + a_{10110} + a_{01110} + a_{00001} - a_{11001} - a_{10101} - a_{01101} + a_{10011} + a_{01011} + a_{00111} - a_{11111})^2} +$$

$$\sqrt{(a_{00000} - a_{11000} - a_{10100} - a_{01100} + a_{10010} + a_{01010} + a_{00110} - a_{11110} + a_{10001} + a_{01001} + a_{00101} - a_{11101} - a_{00011} + a_{11011} + a_{10111} + a_{01111})^2 + (a_{10000} + a_{01000} + a_{00100} - a_{11100} - a_{00010} + a_{11010} + a_{10110} + a_{01110} - a_{00001} + a_{11001} + a_{10101} + a_{01101} - a_{10011} - a_{01011} - a_{00111} + a_{11111})^2} +$$

$$\sqrt{(a_{00000} - a_{11000} + a_{10100} + a_{01100} - a_{10010} - a_{01010} + a_{00110} - a_{11110} - a_{10001} - a_{01001} + a_{00101} - a_{11101} - a_{00011} + a_{11011} - a_{10111} - a_{01111})^2 + (a_{10000} + a_{01000} - a_{00100} + a_{11100} + a_{00010} - a_{11010} + a_{10110} + a_{01110} + a_{00001} - a_{11001} + a_{10101} + a_{01101} - a_{10011} - a_{01011} + a_{00111} - a_{11111})^2} +$$

$$\sqrt{(a_{00000} - a_{11000} + a_{10100} + a_{01100} - a_{10010} - a_{01010} + a_{00110} - a_{11110} + a_{10001} + a_{01001} - a_{00101} + a_{11101} + a_{00011} - a_{11011} + a_{10111} + a_{01111})^2 + (a_{10000} + a_{01000} - a_{00100} + a_{11100} + a_{00010} - a_{11010} + a_{10110} + a_{01110} - a_{00001} + a_{11001} - a_{10101} - a_{01101} + a_{10011} + a_{01011} - a_{00111} + a_{11111})^2} +$$

$$\sqrt{(a_{00000} - a_{11000} + a_{10100} + a_{01100} + a_{10010} + a_{01010} - a_{00110} + a_{11110} - a_{10001} - a_{01001} + a_{00101} - a_{11101} + a_{00011} - a_{11011} + a_{10111} + a_{01111})^2 + (a_{10000} + a_{01000} - a_{00100} + a_{11100} - a_{00010} + a_{11010} - a_{10110} - a_{01110} + a_{00001} - a_{11001} + a_{10101} + a_{01101} + a_{10011} + a_{01011} - a_{00111} + a_{11111})^2} +$$

$$\sqrt{(a_{00000} - a_{11000} + a_{10100} + a_{01100} + a_{10010} + a_{01010} - a_{00110} + a_{11110} + a_{10001} + a_{01001} - a_{00101} + a_{11101} - a_{00011} + a_{11011} - a_{10111} - a_{01111})^2 + (a_{10000} + a_{01000} - a_{00100} + a_{11100} - a_{00010} + a_{11010} - a_{10110} - a_{01110} - a_{00001} + a_{11001} - a_{10101} - a_{01101} - a_{10011} - a_{01011} + a_{00111} - a_{11111})^2} +$$

$$\sqrt{(a_{00000} + a_{11000} - a_{10100} + a_{01100} - a_{10010} + a_{01010} - a_{00110} - a_{11110} - a_{10001} + a_{01001} - a_{00101} - a_{11101} - a_{00011} - a_{11011} + a_{10111} - a_{01111})^2 + (a_{10000} - a_{01000} + a_{00100} + a_{11100} + a_{00010} + a_{11010} - a_{10110} + a_{01110} + a_{00001} + a_{11001} - a_{10101} + a_{01101} - a_{10011} + a_{01011} - a_{00111} - a_{11111})^2} +$$

$$\sqrt{(a_{00000} + a_{11000} - a_{10100} + a_{01100} - a_{10010} + a_{01010} - a_{00110} - a_{11110} + a_{10001} - a_{01001} + a_{00101} + a_{11101} + a_{00011} + a_{11011} - a_{10111} + a_{01111})^2 + (a_{10000} - a_{01000} + a_{00100} + a_{11100} + a_{00010} + a_{11010} - a_{10110} + a_{01110} - a_{00001} - a_{11001} + a_{10101} - a_{01101} + a_{10011} - a_{01011} + a_{00111} + a_{11111})^2} +$$

$$\sqrt{(a_{00000} + a_{11000} - a_{10100} + a_{01100} + a_{10010} - a_{01010} + a_{00110} + a_{11110} - a_{10001} + a_{01001} - a_{00101} - a_{11101} + a_{00011} + a_{11011} - a_{10111} + a_{01111})^2 + (a_{10000} - a_{01000} + a_{00100} + a_{11100} - a_{00010} - a_{11010} + a_{10110} - a_{01110} + a_{00001} + a_{11001} - a_{10101} + a_{01101} + a_{10011} - a_{01011} + a_{00111} + a_{11111})^2} +$$

$$\sqrt{(a_{00000} + a_{11000} - a_{10100} + a_{01100} + a_{10010} - a_{01010} + a_{00110} + a_{11110} + a_{10001} - a_{01001} + a_{00101} + a_{11101} - a_{00011} - a_{11011} + a_{10111} - a_{01111})^2 + (a_{10000} - a_{01000} + a_{00100} + a_{11100} - a_{00010} - a_{11010} + a_{10110} - a_{01110} - a_{00001} - a_{11001} + a_{10101} - a_{01101} - a_{10011} + a_{01011} - a_{00111} - a_{11111})^2} +$$

$$\sqrt{(a_{00000} + a_{11000} + a_{10100} - a_{01100} - a_{10010} + a_{01010} + a_{00110} + a_{11110} - a_{10001} + a_{01001} + a_{00101} + a_{11101} - a_{00011} - a_{11011} - a_{10111} + a_{01111})^2 + (a_{10000} - a_{01000} - a_{00100} - a_{11100} + a_{00010} + a_{11010} + a_{10110} - a_{01110} + a_{00001} + a_{11001} + a_{10101} - a_{01101} - a_{10011} + a_{01011} + a_{00111} + a_{11111})^2} +$$

$$\sqrt{(a_{00000} + a_{11000} + a_{10100} - a_{01100} - a_{10010} + a_{01010} + a_{00110} + a_{11110} + a_{10001} - a_{01001} - a_{00101} - a_{11101} + a_{00011} + a_{11011} + a_{10111} - a_{01111})^2 + (a_{10000} - a_{01000} - a_{00100} - a_{11100} + a_{00010} + a_{11010} + a_{10110} - a_{01110} - a_{00001} - a_{11001} - a_{10101} + a_{01101} + a_{10011} - a_{01011} - a_{00111} - a_{11111})^2} +$$

$$\sqrt{(a_{00000} + a_{11000} + a_{10100} - a_{01100} + a_{10010} - a_{01010} - a_{00110} - a_{11110} - a_{10001} + a_{01001} + a_{00101} + a_{11101} + a_{00011} + a_{11011} + a_{10111} - a_{01111})^2 + (a_{10000} - a_{01000} - a_{00100} - a_{11100} - a_{00010} - a_{11010} - a_{10110} + a_{01110} + a_{00001} + a_{11001} + a_{10101} - a_{01101} + a_{10011} - a_{01011} - a_{00111} - a_{11111})^2} $$

$$\sqrt{(a_{00000} + a_{11000} + a_{10100} - a_{01100} + a_{10010} - a_{01010} - a_{00110} - a_{11110} + a_{10001} - a_{01001} - a_{00101} - a_{11101} - a_{00011} - a_{11011} - a_{10111} + a_{01111})^2 + (a_{10000} - a_{01000} - a_{00100} - a_{11100} - a_{00010} - a_{11010} - a_{10110} + a_{01110} - a_{00001} - a_{11001} - a_{10101} + a_{01101} - a_{10011} + a_{01011} + a_{00111} + a_{11111})^2}\,]^2$$